\documentclass[pra,twocolumn,amsmath,amssymb,showpacs]{revtex4}
\usepackage{amsmath,amssymb,latexsym}
\usepackage{graphicx,epstopdf}

\allowdisplaybreaks[1]

\newcommand{\bs}[1]{\boldsymbol{\it #1}}
\newcommand{\ca}[1]{\mathcal{#1}}

\newcommand{\spaEq}{\hspace{3ex}}
\newcommand{\vspfigA}{\vspace{0em}}  
\newcommand{\vspfigB}{\vspace{0em}} 
\newcommand{\vspfigC}{\vspace{0em}}
\newcommand{\widthfigA}{0.35\textwidth}
\newcommand{\widthfigB}{0.42\textwidth}
\newcommand{\widthfigC}{\widthfigB}

\begin{document}


\title{Escape Dynamics of Many Hard Disks}

\author{Tooru Taniguchi, Hiroki Murata and Shin-ichi Sawada}

\affiliation{School of Science and Technology, 
Kwansei Gakuin University, 2-1 Gakuen, Sanda city, Hyogo, Japan} 

\date{\today}

\begin{abstract} 

   Many-particle effects in escapes of hard disks from a square box via a hole are discussed in a viewpoint of dynamical systems. 
   Starting from $N$ disks in the box at the initial time, we calculate the probability $P_{n}(t)$ for at least $n$ disks to remain inside the box at time $t$  for $n=1,2,\cdots,N$. 
   At early times the probabilities $P_{n}(t)$, $n=2,3,\cdots,N-1$, are described by superpositions of exponential decay functions. 
   On the other hand, after a long time the probability $P_{n}(t)$ shows a power-law decay $\sim t^{-2n}$ for $n\neq 1$, in contrast to the fact that it decays with a different power law $\sim t^{-n}$ for cases without any disk-disk collision.
   Chaotic or non-chaotic properties of the escape systems are discussed by the dynamics of a finite time largest Lyapunov exponent, whose decay properties are related with those of the probability $P_{n}(t)$. 

\end{abstract}


\vspace{0.5cm} 
\maketitle

\section{Introduction}

   The escape of materials from a finite area is known as an essential concept to understand many physical features in a variety of natural phenomena. 
   It describes a wide scale of physical phenomena from a microscopic scale (e.g. $\alpha$ decay of nuclei \cite{G28,GC29} and light emissions from molecules \cite{RLK06,SH07,K10}) to a macroscopic one (e.g. ejections or evaporations of stars in cosmology \cite{N04,BT87}). 
   It also plays an important role to analyze properties of materials, for example, by means of particle escapes from a quantum dot \cite{TM87,CS00,MN08} or from an optical potential trap \cite{WB97,FG01,SZ11}, escape basins of magnetic field lines in plasma \cite{EM02,VS11}, and transition of states in chemical reactions (as an escape of excited chemical species from a reactant region) \cite{K40,T07,CK12}, etc. 
   The concept of escape is also used to calculate transport coefficients in chaotic dynamical systems \cite{GN90,G98,D99,K07}, and as a mechanism to produce electric currents as escapes of electrons from particle reservoirs \cite{D95,I97,D98}. 
   Escapes occur when particles reach a specific region like a hole, so that they are related to the first passage time problem and the recurrence problem \cite{R89,K92,AT09}. 
   In escape phenomena, states of materials leave their initial ones with a decay, and this  type of dynamical features also appears in the Loschmidt echo and fidelity decay \cite{GP06,PP09,GJP12}. 

   Escape phenomena are characterized by various quantities, such as the survival probability  \cite{BB90,ZB03,BD07,DG09,AP13,TS11a,TS11b,TS13} as the probability for a particle to remain in an initially confined area, the escape time \cite{AT09,AGH96,PB00} as the time period for a particle to stay in the initial area, and a velocity of a particle escaping from the initial area \cite{TS11b}, etc. 
   These quantities decay in time as a feature of escape phenomena in which materials continue leaking from the initial area. 
   Many works have already been done to clarify their decay properties in escapes of a single particle by using dynamical theories. 
   It is conjectured, for example, that the survival probability decays exponentially in time for chaotic systems based on an ergodic argument, while it decays with a power law for non-chaotic systems \cite{BB90}. 
   This conjecture led to further dynamical studies of escape phenomena, clarifying effects of a finite size of holes \cite{BD07,AGH96}, weakness of chaos \cite{AT09}, specific orbits causing a power-law decay \cite{AH04,DG09}, etc.

   The principal aim of this paper is to discuss many-particle effects in dynamical properties of escape phenomena. 
   As a system consisting of many particles, we consider many hard disks in a square box, which have been widely used to investigate statistical and dynamical properties \cite{Del96,S00,TM05}.  
   At a corner of the box we put a hole where hard disks escape from the box. 
   To discuss escape properties of $N$ disks from the box via the hole, we introduce the survival probability $P_{n}(t)$ for $n$ disks to remain inside the box at time $t$ ($n=1,2,\cdots,N$). 
   We show that at early times decays of the survival probabilities $P_{n}(t)$, $n=2,3,\cdots,N-1$, are well-described by superpositions of exponential functions. 
   It is also shown that after a long time the survival probability $P_{n}(t)$ shows a power-law decay $\sim t^{-2n}$, while it decays with a different power law $\sim t^{-n}$ for the case without any disk-disk collision, for $n=2,3,\cdots$.
   These results mean qualitative changes in decays of survival probabilities by disk-disk collisions and changing numbers of disks, implying a possibility to get information on particle-particle interactions and the number of non-escaping particles from their decay behaviors.  

   In this paper we also discuss dynamical properties of escape systems by using the finite time largest Lyapunov exponent (FTLLE) $\lambda (t)$ \cite{O93,CC10,ST13}, which is introduced as an exponential rate of expansion or contraction of an infinitesimally small initial error at a finite time $t$. 
   The FTLLE converges to the well-known largest Lyapunov exponent in the long time limit, whose positivity means dynamics of the system to be chaotic. 
   We compare quantitatively decay properties of the survival probability $P_{n}(t)$ and the corresponding FTLLE $\lambda (t)$, and clarify roles of chaotic or non-chaotic dynamics in escape phenomena of many hard disks. 
  Dependences of the survival probabilities and the FTLLEs on the system length and the hole size, etc., are also discussed as scaling properties. 
   
   The outline of this paper is as follows. 
   In Sec. \ref{EscapesChaoticManyHardDiskSystems} we introduce our model consisting of many hard disks in a square box with a hole, and discuss exponential and power-law decays of survival probabilities of this system. 
   In Sec. \ref{FiniteTimeLyapunovExponentsEscapeEystems} we discuss decay properties of FTLLEs in escape systems with many hard disks, and investigate connections between decay properties of the survival probabilities and the FTLLEs. 
   Finally, we give conclusions and remarks on the contents of this paper in Sec. \ref{ConclusionRemarks}.

\section{Escape properties of many-hard-disk systems} 
\label{EscapesChaoticManyHardDiskSystems}

\subsection{Many hard disks in a square box with a hole}

\begin{figure}[!t]
\vspfigA
\begin{center}
\includegraphics[width=\widthfigA]{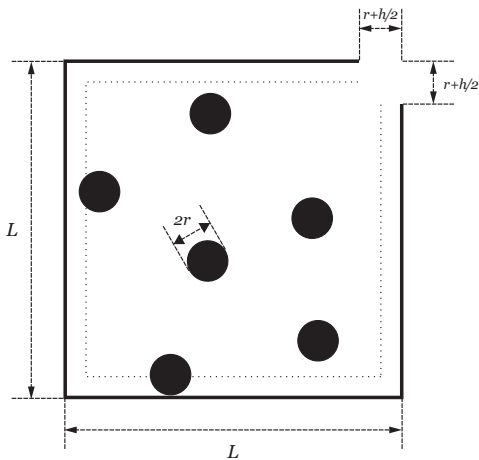}
\vspfigB
\caption{
   Many hard disks with the radius $r$ in a square box with the length $L$ of each side. 
   The box has a hole with each length $r+(h/2)$ of both the sides from a corner of the box. 
   The dotted lines in inner sides of the box are for positions of the center of a disk closest to a wall of the box.   
}
\label{Fig1SystemDisks}
\end{center}
\vspfigC
\end{figure}

   We consider the system consisting of many hard disks inside a square box with a hole. 
   Here, the mass and the radius of the disks are $m$ and $r$, respectively, and the length of each side of the box is $L$, and a hole in the box is taken as the region of the length $r+(h/2)$ in both the sides from a single corner of the box, as shown in Fig. \ref{Fig1SystemDisks} as a schematic illustration. 
   Here, the length $h/2$ is the effective length of each side of the hole, where centers of disks can reach.  
   In this system, movements of each disk consists of uniform linear motions, elastic collisions with other disks or walls, and an escape from the box via the hole. 
   Besides, it is assumed that any disk does not enter into the box via the hole from its outside, and any disk is removed from the box when it reaches the hole. 
   As an important system parameter we use the particle density $\rho \equiv N \pi r^{2}/L^{2}$ of the system at the initial time $t=0$. 
%

\subsection{Survival probabilities of many-hard-disk systems}

   To characterize escape behaviors of $N$ disks from a square box via a hole, we introduce the probability $P_{n}(t)$, for which at least $n$ disks remain inside the box at time $t$, for $n=1,2,\cdots,N$ \cite{NoteB}. 
   We call this probability $P_{n}(t)$ "the $n$-particle survival probability", or simply the survival probability, as a generalization of the well-known survival probability discussed in one-particle escape systems \cite{DG09,BB90,AP13}. 
   Starting from almost arbitrary initial conditions, except for some specific cases, for example, that disks move in a periodic orbit without reaching the hole, the $n$-particle survival probability goes to zero, i.e., $ \lim_{t\rightarrow +\infty} P_{n} (t) = 0$, in the long time limit $t \rightarrow +\infty$, and escape properties of many-particle systems are characterized by decay behaviors of $P_{n}(t)$.

\begin{figure}[!t] 
\vspfigA
\begin{center}
\includegraphics[width=\widthfigB]{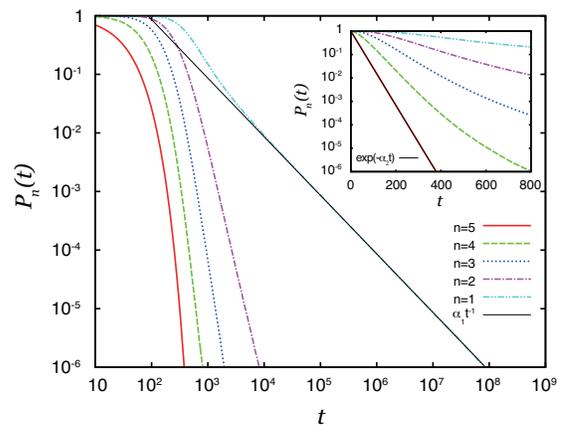}
\vspfigB
\caption{
   (Color online) 
   The $n$-particle survival probabilities $P_{n}(t)$ of a $5$-disk system as functions of time $t$  for $n = 5$ (the solid line), $4$ (the dashed line), $3$ (the dotted line), $2$ (the dash-dotted line), $1$ (the dash-double-dotted line). 
   The main figure is for graphs of $P_{n}(t)$ on log-log plots with a straight shin line for a fitting function $\alpha_{1}t^{-1}$ to $P_{1}(t)$ with the fitting parameters $\alpha_{1}$. 
   The inset is for graphs of $P_{n}(t)$ on linear-log plots with a straight shin line for a fitting function $\exp (-\alpha_{2}t)$ to $P_{5}(t)$ with the fitting parameter $\alpha_{2}$.
}
\label{Fig2NParSurProb}
\end{center}
\vspfigC
\end{figure}  
%
   In Fig. \ref{Fig2NParSurProb} we show graphs of the $n$-particle survival probabilities $P_{n}(t)$ of $N$ hard disks as functions of time $t$ for $n = 5$ (the solid line), $4$ (the dashed line), $3$ (the dotted line), $2$ (the dash-dotted line), $1$ (the dash-double-dotted line) with $N=5$. 
   Here, we used values of the parameters as $m=1$, $r=1/2$, $h = 0.1(L-2r)$ and $\rho = 0.1$ (so $L \approx 6.27$). 
   For the survival probabilities shown in Fig. \ref{Fig2NParSurProb}, we calculated $5\times 10^{8}$ number of ensembles from random initial conditions at $t=0$ in which the disk positions and momenta are distributed into the microcanonical distribution of the system without the hole with the value $N$ of energy $E$ \cite{SF78,D91}.

   In the main figure of Fig. \ref{Fig2NParSurProb}, we show the log-log plots of the $n$-particle survival probabilities $P_{n}(t)$, $n=1,2,\cdots,5$. 
   As shown clearly as a straight line in this figure, the $1$-particle survival probability $P_{1}(t)$ shows a power-law decay after a long time. 
   We fitted this power-law decay of $P_{1}(t)$ to a power function $\alpha_{1}t^{-1}$ with the value $\alpha_{1}= 86.6$ of the fitting parameter $\alpha_{1}$.
   This result is simply explained by the fact that only a single disk exists in the majority of times in the decay of $P_{1}(t)$ and a single disk in the square box is not chaotic, leading to the power-law decay $\sim t^{-1}$ of the survival probability \cite{BB90}. 
%
   On the other hand, as shown in the inset of Fig. \ref{Fig2NParSurProb} as linear-log plots of $P_{n}(t)$, the $N$-particle survival probability $P_{N}(t)$ decays exponentially in time. 
   To clarify this property, we fitted the graph of $P_{N}(t)$ to an exponential function $\exp(-\alpha_{2} t)$ with the value $\alpha_{2} = 3.65\times 10^{-2}$ of the fitting parameter $\alpha_{2}$, which is almost indistinguishable with the graph of $P_{N}(t)$ in the inset of Fig. \ref{Fig2NParSurProb}. 
   It should be noted that an exponential decay of the survival probability also appears in one-particle chaotic systems \cite{BB90}.  

\subsection{Exponential decays of survival probabilities}
\label{ExponentialDecaysSurvivalProbabilities}

   Now, we proceed to discuss decay properties of the $n$-particle survival probabilities $P_{n}(t)$ for the middle numbers $n=2,3,\cdots,N-1$. 
   The inset of Fig. \ref{Fig2NParSurProb} suggests that different from the graph of $P_{N}(t)$, the graphs of $P_{n}(t)$ for $n=2,3,\cdots,N-1$ do not seem to show simple exponential decays at early times, although the dynamics of $n$ disks in the square box for $n=2,3,\cdots,N-1$ can be chaotic with disk-disk collisions.

   In order to describe decay behaviors of the survival probabilities $P_{n}(t)$, $n=2,3,\cdots,N-1$, at early times, we introduce the probability density $f_{k}(\tau)$ of the time $\tau$ for the $k$-th disk escape to occur (i.e., the escape time $\tau$ of $k$ disks), for $k=1,2,\cdots,N$. 
   Using the escape-time probability density $f_{k}(\tau)$ of $k$ disks, the $(N-k+1)$-particle survival probability $P_{N-k+1}(t)$ for $N-k+1$ disks to remain inside the box at time $t$ is represented as 
\begin{eqnarray}
   P_{N-k+1}(t) = 1-\int_{0}^{t}d\tau\; f_{k}(\tau) ,
\label{EscapTimeProba1}
\end{eqnarray}
so that we obtain 
\begin{eqnarray}
   f_{k}(\tau) = -\frac{d P_{N-k+1}(\tau)}{d\tau} ,
\label{EscapTimeProba2}
\end{eqnarray}
$k=1,2,\cdots,N$. 
   By Eq. (\ref{EscapTimeProba2}), for example, if the $N$-particle survival probability $P_{N}(t)$ decays exponentially in time, i.e., $P_{N}(t) = \exp(-at)$ with a positive constant $a$ as shown in Fig. \ref{Fig2NParSurProb}, then we obtain the escape-time probability density $f_{1}(\tau) = \tilde{f}_{1}(\tau;a)$ for the first escaping disk, in which $\tilde{f}_{1}(\tau;a)$ is given by 
\begin{eqnarray}
   \tilde{f}_{1}(\tau;a) = a e^{-a\tau}
\label{FunctTildeF1}
\end{eqnarray}
for $N>1$.
   Equation (\ref{EscapTimeProba1}) or (\ref{EscapTimeProba2}) also leads to the normalization condition $\int_{0}^{+\infty}d\tau\;f_{k}(\tau) = 1$ of the escape-time probability density $f_{k}(\tau)$ as far as the survival probability $P_{N-k+1}(t)$ goes to zero in the long time limit, i.e., $\lim_{t\rightarrow +\infty} P_{N-k+1}(t) = 0$, noting the initial condition $P_{N-k+1}(0) = 1$ of the survival probability.

   The escape time $\tau$ of $k$ disks is represented as the sum $\tau = \sum_{j=1}^{k} \tau_{j}$ of the time-interval $\tau_{j} \equiv t_{j}-t_{j-1}$ from the time $t_{j-1}$ of the $(j-1)$-th escape to the time $t_{j}$ of the $j$-th escape of the $k$ disks, defining $t_{0}\equiv 0$. 
   For the case that the dynamics of $k$ disks inside the box is chaotic and the probability for each disk to stay inside the box decays exponentially in time, we assume that the probability density of the time $\tau_{k}$ for the $k$-th escaping disk is independent of other times $\tau_{j}$, $j = k-1,k-2,\cdots,1$, and the probability density $f_{k}(\tau)$ of the escape time $\tau$ satisfies the recurrence relation
%
$
   f_{k}(\tau) = \int_{0}^{\tau}ds\; f_{1}(\tau-s)f_{k-1}(s) 
$
%
with $f_{1}(\tau_{k}) = \tilde{f}_{1}(\tau_{k};a_{k})$ and a positive constant $a_{k}$. 
   Under this assumption, we obtain the probability density $f_{k}(\tau) = \tilde{f}_{k}(\tau)$ of the escape time $\tau$ of the $k$ disks with $k\geq 2$, in which $\tilde{f}_{k}(\tau)$ is given by
\begin{widetext}
\begin{eqnarray}
   \tilde{f}_{k}(\tau) &=& \int_{0}^{\tau}dt_{k}\tilde{f}_{1}(\tau-t_{k};a_{k})
      \int_{0}^{t_{k}}dt_{k-1}\tilde{f}_{1}(t_{k}-t_{k-1};a_{k-1})
      \nonumber \\
   &&\spaEq\times
      \int_{0}^{t_{k-1}}dt_{k-2}\tilde{f}_{1}(t_{k-1}-t_{k-2};a_{k-2})
      \cdots 
      \int_{0}^{t_{3}}dt_{2}\tilde{f}_{1}(t_{3}-t_{2};a_{2})
      \tilde{f}_{1}(t_{2};a_{1}) 
\label{EscapTimeProba3a}
       \\
   &=& \left(\prod_{j=1}^{k} a_{j}\right)\sum_{j=1}^{k} \frac{e^{-a_{j} \tau}}{(a_{1}-a_{j})(a_{2}-a_{j})\cdots (a_{j-1}-a_{j})(a_{j+1}-a_{j})(a_{j+2}-a_{j})\cdots (a_{k}-a_{j})} ,
\label{EscapTimeProba3b}
\end{eqnarray}
in which we assumed the condition $a_{j} \neq a_{k}$ for $j\neq k$. 
   From Eqs. (\ref{EscapTimeProba1}) and (\ref{EscapTimeProba3b}) we derive the $(N-k+1)$-particle survival probability $P_{N-k+1}(t) = \tilde{P}_{N-k+1}(t)$, in which $\tilde{P}_{N-k+1}(t)$ is given by
\begin{eqnarray}
   \tilde{P}_{N-k+1}(t) = \sum_{j=1}^{k} \frac{a_{1}a_{2}\cdots a_{j-1}a_{j+1}a_{j+2}\cdots a_{k} 
   e^{-a_{j} t}}{(a_{1}-a_{j})(a_{2}-a_{j})\cdots (a_{j-1}-a_{j})(a_{j+1}-a_{j})(a_{j+2}-a_{j})\cdots (a_{k}-a_{j})}
\label{SurviProbaExpon1}
\end{eqnarray}
\end{widetext}
for $k=2,\cdots,N-1$.
   The derivation of Eq. (\ref{EscapTimeProba3b}) from Eqs. (\ref{FunctTildeF1}) and (\ref{EscapTimeProba3a}), as well as the derivation of Eq. (\ref{SurviProbaExpon1}) from Eqs. (\ref{EscapTimeProba1}) and (\ref{EscapTimeProba3b}), are given in Appendix \ref{LaplaceHeaviside}.
   Here, the $1$-particle survival probability $P_{1}(t)$ would not be justified by Eq. (\ref{SurviProbaExpon1}), because the dynamics of the last single disk inside the box is not chaotic.

\begin{figure}[!t] 
\vspfigA
\begin{center}
\includegraphics[width=\widthfigB]{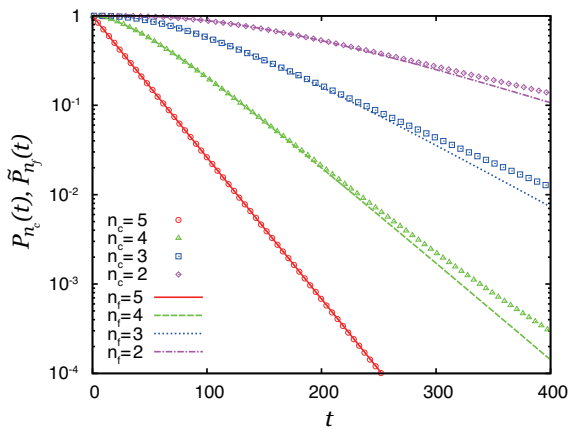}
\vspfigB
\caption{
   (Color online)
   Fittings of the $n$-particle survival probabilities $P_{n}(t)$ of a 5-disk system to the functions $\tilde{P}_{n}(t)$ as functions of time $t$ for $n = 5,4,3,2$ with the the fitting parameters $a_{j}$, $j=1,2,3,4$, on linear-log plots. 
   Here, $P_{n}(t)$ and $\tilde{P}_{n}(t)$ are shown as the open circles and the solid line for $n=5$, the triangles and the broken line for $n=4$, the squares and the dotted line for $n=3$, and the diamonds and the dash-dotted line for $n=2$, respectively.
}
\label{Fig3SurProExp}
\end{center}
\vspfigC
\end{figure}  
%
   In Fig. \ref{Fig3SurProExp} we show fittings of the $n$-particle survival probabilities $P_{n}(t)$ for $n=5$ (the circles), $4$ (the triangles), $3$ (the squares) and $2$ (the diamonds) to the functions $\tilde{P}_{n}(t)$ for $n=5$ (the solid line), $4$ (the broken line), $3$ (the dotted line), $2$ (the dash-dotted line), respectively, with fitting parameters $a_{j}$, $j=1,2,\cdots,4$. 
   Here, the $n$-particle survival probabilities $P_{n}(t)$ in this figure are the same as those in Fig. \ref{Fig2NParSurProb}, and we defined the function $\tilde{P}_{N}(t)$ by $\tilde{P}_{N}(t) \equiv \exp(-a_{1}t)$, while the functions $\tilde{P}_{n}(t)$, $n=2,\cdots,N-1$, are given by Eq. (\ref{SurviProbaExpon1}). 
   In these fittings, we first fitted the $N$-particle survival probability $P_{N}(t)$ to $\tilde{P}_{N}(t)$ with a fitting parameters $a_{1}$, then fitted the $(N-j+1)$-particle survival probability $P_{N-j+1}(t)$ to $\tilde{P}_{N-j+1}(t)$ with a fitting parameters $a_{j}$ and the already fitted values of $a_{1}, a_{2}, \cdots, a_{j-1}$, for $j=2,3,\cdots,N-1$.   
   The values of the fitting parameters $a_{j}$, $j=1,2,\cdots,4$ used for the functions $\tilde{P}_{n}(t)$ shown in Fig. \ref{Fig3SurProExp} are given as the data for $\rho = 10^{-1}$ and $h/(L-2r) = 10^{-1}$ in Table \ref{ExponDEcayFitting}. 
   As shown in Fig. \ref{Fig3SurProExp}, the function (\ref{SurviProbaExpon1}) fits the $n$-particle  survival probability $P_{n}(t)$ reasonably well at early times for $n = 2,3,\cdots,N-1$.

\newcommand{\spaTabA}{\hspace{1.8ex}}
\newcommand{\spaTabB}{\spaTabA\spaTabA}
\begin{table}[!t] 
\vspfigA
\begin{center}
\begin{tabular}{c|c|c|c||c|c|c|c}
  $\rho$ & $ L\!-\!2r $ & $h/(L\!-\!2r)$ & $\tilde{a}$ &
      $a_{1}/\tilde{a}$  &  $a_{2}/\tilde{a}$  & 
      $a_{3}/\tilde{a}$  &  $a_{4}/\tilde{a}$  
     \\ \hline
  $10^{-1}$ & $5.27$ & $10^{-1}$ & $2.69\times 10^{-2}$ &
     $1.36$ & $0.933$ & $0.607$ & $0.358$ 
     \\ \hline
  $10^{-2}$ & $18.8$ & $10^{-1}$ &  $7.52\times 10^{-3}$ &
     $1.08$ & $0.802$ & $0.565$ & $0.357$ 
     \\ \hline
  $10^{-3}$ & $61.7$ & $10^{-1}$ &  $2.29\times 10^{-3}$ &
     $1.08$ & $0.806$ & $0.578$ & $0.355$
     \\ \hline
  $10^{-1}$ & $5.27$ & $10^{-2}$ &  $2.69\times 10^{-3}$ &
     $1.37$ & $0.952$ & $0.632$ & $0.369$
     \\ \hline
  $10^{-1}$ & $5.27$ & $10^{-3}$ &  $2.69\times 10^{-4}$ &
     $1.38$ & $0.956$ & $0.640$ & $0.372$
     \\ \hline
\end{tabular}
\vspfigB
\caption{Decay rates $a_{n}$, $n=1,2,3,4$, divided by $\tilde{a}\equiv hv_{0}/(L-2r)^{2}$, for a $5$-disk system with various values of the initial particle density $\rho$ and the hole size ratio $h/(L-2r)$.}
\label{ExponDEcayFitting}
\end{center}
\vspfigC
\end{table}
%
   In Table \ref{ExponDEcayFitting} we show the fitting values of the exponential decay rates $a_{j}$, $j = 1,2,3,4$, divided by the quantity $\tilde{a} \equiv hv_{0}/(L-2r)^{2}$ with the one-particle initial average speed $v_{0}\equiv \sqrt{2E/(mN)}$, for various initial particle densities $\rho$ and ratios $h/(L-2r)$ of the hole size $h$ to the effective side length $L-2r$ of the box, in the survival probabilities $P_{n}(t)$, $n = 1,2,3,4$ of $5$-disk systems at early times.   
   Here, the system parameters used to obtain these data, other than those shown in Table \ref{ExponDEcayFitting} and the ensemble number $2\times 10^{8}$ only for the case of $(\rho,h/(L-2r)) = (10^{-1},10^{-3})$, are the same as those of the system whose survival probabilities are shown in Fig. \ref{Fig2NParSurProb}.  
   The factor $1/\tilde{a}$ used for the quantities $a_{j}/\tilde{a}$ in this table, comes from the fact that the exponential decay rate of the survival probability in escapes of a single chaotic point particle via a small hole in a two-dimensional space is proportional to $hv_{0}/\ca{S}$ with the hole size $h$, the particle speed $v_{0}$, and the area $\ca{S}$ for the point particle to move before escaping \cite{BB90}. 
   Table \ref{ExponDEcayFitting} suggests that the quantity $a_{j}$/$\tilde{a}$ for each value of $j$ takes similar values for a wide variety of initial particle densities $\rho$ and hole size ratios $h/(L-2r)$ like in escapes of a single particle, although the value of $a_{j}$/$\tilde{a}$ decreases as the index $j$ increases.

\subsection{Disk-disk collisions and power-law decays of survival probabilities} 
\label{PowerDecaysSurvivalProbabilities} 

   Now, we discuss effects of disk-disk collisions in decays of the $n$-particle survival probability $P_{n}(t)$ after a long time. 
   To clarify such effects, we compare decay features of survival probabilities in two different types of many-particle systems: the many-hard-disk systems with disk-disk collisions, and the systems consisting of many disks which can overlaps with each other, i.e., without any disk-disk collision.  
   Here, the disk systems without any disk-disk collision may be regarded as the systems consisting of $N$ point-particles in the square box which has the side length $L-2r$ and the hole in the region of the length $r$ in both the sides from a single corner of the box.

\begin{figure}[!t] 
\vspfigA
\begin{center}
\includegraphics[width=\widthfigB]{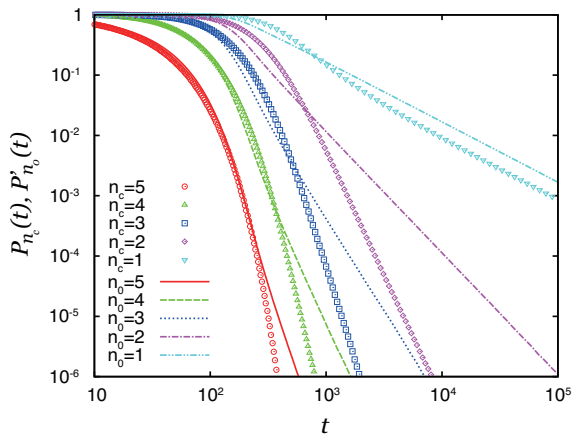}
\vspfigB
\caption{
   (Color online)
   The $n_{0}$-particle survival probabilities $P_{n_{0}}^{\prime}(t)$ of a 5-disk system without any disk-disk collision as functions of time $t$ for $n_{0} = 5$ (the solid line), $4$ (the dashed line), $3$ (the dotted line), $2$ (the dash-dotted line) and $1$ (the dash-double-dotted line), and the $n_{c}$-particle survival probability $P_{n_{c}}(t)$ of a 5-disk system with disk-disk collisions as functions of time $t$ for $n_{c}=5$ (the circles), $4$ (the triangles), $3$ (the squares), $2$ (the diamonds) and $1$ (the inverted triangles),  on log-log plots. 
}
\label{Fig4SurProbComp}
\end{center}
\vspfigC
\end{figure}  
%
   Figure \ref{Fig4SurProbComp} is the graphs of the $n_{0}$-particle survival probabilities $P_{n_{0}}^{\prime}(t)$ of 5 disks without any disk-disk collision for $n_{0} = 5$ (the solid line), $4$ (the dashed line), $3$ (the dotted line), $2$ (the dash-dotted line) and $1$ (the dash-double-dotted line), as well as the $n_{c}$-particle survival probabilities $P_{n_{c}}(t)$ of 5 disks with their collisions for $n_{c} = 5$ (the circles), $4$ (the triangles), $3$ (the squares), $2$ (the diamonds) and $1$ (the inverted triangles). 
   Here, the survival probabilities $P_{n_{c}}(t)$ are the same as in Fig. \ref{Fig2NParSurProb}, and all values of the system parameters of the system for the survival probabilities $P_{n_{0}}^{\prime}(t)$ are the same as those for $P_{n_{c}}(t)$ except for the conditions in which disks can overlap without any impact in their time-evolutions and the initial velocity distribution of each particle is given by a uniform distribution under the constraint for the speed of each particle to be $1$, imposing the microcanonical distribution for each independent particle at the initial time $t=0$.

   It is shown in Fig. \ref{Fig4SurProbComp} that at some early times the survival probability $P_{n}^{\prime}(t)$ of the system without any disk-disk collision can decay faster than the corresponding survival probability $P_{n}(t)$ of the system with disk-disk collisions, for $n = 4,3,\cdots,1$. 
   In contrast, after a long time the survival probability $P_{n}^{\prime}(t)$ decays slower than the corresponding survival probability $P_{n}(t)$, for $n = 5,4,\cdots,1$. 
   We can also recognize in Fig. \ref{Fig4SurProbComp} that after a long time the survival probability $P_{1}^{\prime}(t)$ shows a power-law decay $\sim t^{-1}$, similar to the survival probability $P_{1}(t)$.

\begin{figure}[!t] 
\vspfigA
\begin{center}
\includegraphics[width=\widthfigC]{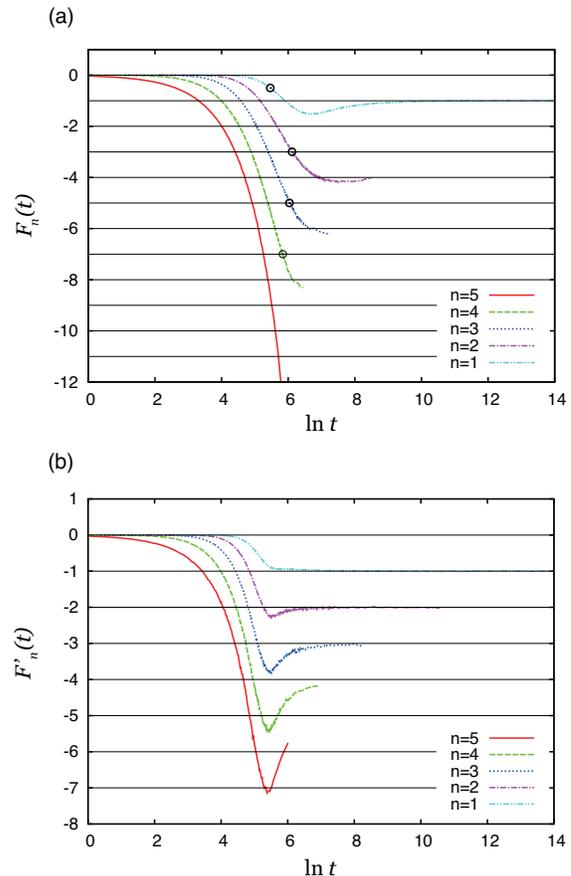}
\vspfigB
\caption{
   (Color online)
   (a) The slopes $F_{n}(t)$ of $\ln P_{n}(t)$, and (b) the slopes $F_{n}^{\prime}(t)$ of $\ln P_{n}^{\prime}(t)$, as functions of $\ln t$ for $n = 5$ (the solid line), $4$ (the dashed line), $3$ (the dotted line), $2$ (the dash-dotted line) and $1$ (the dash-double-dotted line)  for $5$-disk systems.  
   The open circles in (a) are the points for the slope $F_{n}(t)$ to cross the line $-(2n-1)$ for $n=4,3$ and $2$, and the line $-1/2$ for $n=1$.
}
\label{Fig5SlopSurProb}
\end{center}
\vspfigC
\end{figure}  
%
   It is clearly shown in Fig. \ref{Fig4SurProbComp} that unlike the survival probability $P_{1}^{\prime}(t)$, a decay behavior of the survival probability $P_{n}^{\prime}(t)$ is essentially different from that of the corresponding survival probability $P_{n}(t)$ for $n=2,3,\cdots,N-1$ after a long time. 
   On the other hand, it is not clear by the fittings of $P_{n}^{\prime}(t)$, $n=2,3,\cdots,N-1$, to power functions in Fig. \ref{Fig4SurProbComp} whether these survival probabilities $P_{n}^{\prime}(t)$ exhibit asymptotic power-law decays in time or not. 
   To clarify this point quantitatively, we introduce the function $F_{n}(t)$ defined by 
\begin{eqnarray}
   F_{n}(t) \equiv \frac{d \ln P_{n}(t)}{d \ln t} = \frac{t}{P_{n}(t)}\frac{d P_{n}(t)}{d t} ,
\label{SlopSurProb1}
\end{eqnarray}
which takes a constant value $-\nu$ of the power in the case that the $n$-particle survival probability $P_{n}(t)$ shows a power-law decay $\sim t^{-\nu}$ \cite{NoteC}. 
   In Fig. \ref{Fig5SlopSurProb}(a) we show the slopes $F_{n}(t)$ of $\ln P_{n}(t)$ as functions of $\ln t$  for $n = 5$ (the solid line), $4$ (the dashed line), $3$ (the dotted line), $2$ (the dash-dotted line) and $1$ (the dash-double-dotted line), for the $5$-disk system with disk-disk collisions. 
   (It may be noted that if the survival probability $P_{n}(t)$ decays exponentially in time as $P_{n}(t) = \exp (-at)$ with a positive constant $a$, then the function (\ref{SlopSurProb1}) becomes $-at$, leading to the initial value $F_{n}(0) = 0$ of the function $F_{n}(t)$ as shown in Fig. \ref{Fig5SlopSurProb}(a).)
   For a comparison, in Fig. \ref{Fig5SlopSurProb}(b) we also show the slopes $F_{n}^{\prime}(t)\equiv d \ln P_{n}^{\prime}(t)/d \ln t$ of $\ln P_{n}^{\prime}(t)$ as functions of $\ln t$  for $n = 5$ (the solid line), $4$ (the dashed line), $3$ (the dotted line), $2$ (the dash-dotted line) and $1$ (the dash-double-dotted line), for the corresponding $5$-disk system  without any disk-disk collision. 
   Here, we used the systems whose $n$-particle survival probabilities $P_{n}(t)$ and $P_{n}^{\prime}(t)$ are shown in Fig. \ref{Fig4SurProbComp}.

   Figure \ref{Fig5SlopSurProb} shows that both the $1$-particle survival probabilities $P_{1}(t)$ and $P_{1}^{\prime}(t)$ exhibit a power-law decay $\sim t^{-1}$ after a long time, but power-law decay properties of other $n$-particle survival probabilities $P_{n}(t)$ and $P_{n}^{\prime}(t)$ for $n=2,3,\cdots,N-1$ after a long time are different with each other. 
   For the case with disk-disk collisions, Fig. \ref{Fig5SlopSurProb}(a) shows that the survival probabilities $P_{2}(t)$ and $P_{3}(t)$ exhibit a  power-law decay $\sim t^{-4}$ and $\sim t^{-6}$, respectively, suggesting their asymptotic power-law decays as 
\begin{eqnarray}
   P_{n}(t) \;\overset{t\rightarrow+\infty}{\sim}\; 
      \left\{\begin{array}{ll}
      \eta_{1}t^{-1}  &   \;\;\;\mbox{for}\; n=1   \\
      \eta_{n} t^{-2n} &   \;\;\;\mbox{for}\; n=2,3,\cdots ,
   \end{array}\right.
\label{PowerDecaySurProbCol1}
\end{eqnarray}
with constants $\eta_{n}$, $n=1,2,\cdots$, although the graph of $F_{4}(t)$ in Fig. \ref{Fig5SlopSurProb}(a) does not show clearly such an asymptotic power-law decay of the survival probability $P_{4}(t)$ yet. 
   In contrast, for the case without any disk-disk collision, Fig. \ref{Fig5SlopSurProb}(b) shows that the survival probabilities  $P_{2}^{\prime}(t)$, $P_{3}^{\prime}(t)$ and $P_{4}^{\prime}(t)$ decays with a power law $\sim t^{-2}$, $\sim t^{-3}$ and $\sim t^{-4}$, respectively, suggesting their asymptotic decays as 
\begin{eqnarray}
   P_{n}^{\prime}(t) \;\overset{t\rightarrow+\infty}{\sim}\; 
      \eta_{n}^{\prime} t^{-n}   
\label{PowerDecaySurProbNoCol1}
\end{eqnarray}
with a constant $\eta_{n}^{\prime}$ for $n=1,2,\cdots$.
   The difference between Eqs. (\ref{PowerDecaySurProbCol1}) and (\ref{PowerDecaySurProbNoCol1}) would be regarded as an important effect of disk-disk collisions in decays of the $n$-particle survival probabilities.

\section{Finite-time Lyapunov exponents of escape systems}   
\label{FiniteTimeLyapunovExponentsEscapeEystems}
\subsection{Decays of finite-time Lyapunov exponents}
\label{DecaysFiniteTimeLyapunovExponents}

   The many-hard-disk system considered in this paper is chaotic, as far as two colliding hard disks exist inside the box.  
   In order to characterize chaotic dynamics of the system with a dynamical instability, we introduce the finite-time largest Lyapunov exponent (FTLLE) $\lambda (t)$ at time $t$, which is defined by 
\begin{eqnarray}
   \lambda (t) = \lim_{|\delta \bs{\Gamma}(0)|\rightarrow 0}\frac{1}{t}\ln \frac{ |\delta \bs{\Gamma}(t) |}{|\delta \bs{\Gamma}(0)|} .
\label{FTLLE1}
\end{eqnarray}
Here, $\delta \bs{\Gamma}(t)$ is a small deviation of the phase space vector (consisting of the position vector and the momentum vector) of the hard disks inside the box at time $t$, and the dimension of the vector $\delta \bs{\Gamma}(t)$ reduces by four at every time when a disk escapes from the hole. 
   One may notice that using the discretized times $t_{k} = k\Delta t$, $k = 0,1,\cdots, K$ with $\Delta t \equiv t/K$ and an integer $K$, Eq. (\ref{FTLLE1}) is rewritten as 
\begin{eqnarray}
   \lambda (t) = \lim_{|\delta \bs{\Gamma}(0)|\rightarrow 0}\frac{1}{K} \sum_{k=1}^{K} \frac{1}{\Delta t}\ln \frac{ |\delta \bs{\Gamma}(t_{k}) |}{|\delta \bs{\Gamma}(t_{k-1})|} 
\end{eqnarray}
meaning that the FTLLE  $ \lambda (t)$ is given from an average of the quantities $(1/\Delta t)\ln (|\delta \bs{\Gamma}(t_{k}) |/|\delta \bs{\Gamma}(t_{k-1})|) $ indicating local time dynamical instabilities in the limit of $K \rightarrow +\infty$, i.e., $\Delta t \rightarrow +0$, characterizing a majority of such dynamical properties in the finite time interval $[0,t]$. 
   The well-known largest Lyapunov exponent of the system in the non-escape case with $h=0$ is given by $\lim_{t\rightarrow +\infty}\lambda (t)$. 
   On the other hand, in the case for disks to escape with $h > 0$, even if the system is chaotic initially, then we have $\lim_{t\rightarrow +\infty}\lambda (t) = 0$ for almost any initial condition for which no disk exists inside the box in the long time limit $t\rightarrow +\infty$. 
   Therefore, we could characterize chaotic or non-chaotic properties of the system by a decay behavior of the FTLLE $\lambda (t)$ at time $t$.

\begin{figure}[!t]
\vspfigA
\begin{center}
\includegraphics[width=\widthfigB]{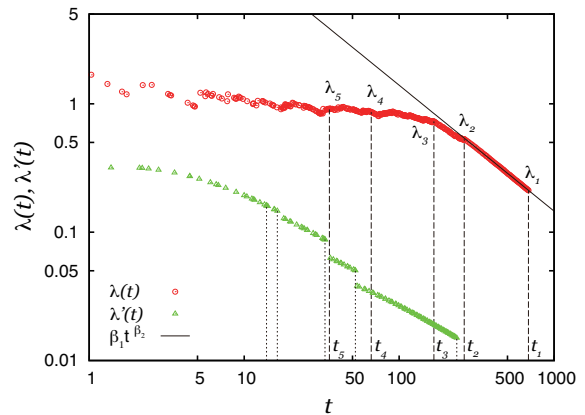}
\vspfigB
\caption{
   (Color online)
   The finite-time largest Lyapunov exponents (FTLLEs) as a function of time $t$ on log-log plots; $\lambda (t)$ (circles) for a 5-disk system with disk-disk collisions, and $\lambda^{\prime}(t)$ (triangles) for one without any disk-disk collision. 
   Here, the broken vertical lines and the dotted ones indicate the times $t_{n}$ when the $n$-th escape of the disks from the box occurs with the value $\lambda_{n}$ of the FTLLEs $\lambda (t)$ and $\lambda^{\prime}(t)$, respectively, for $n=1,2,\cdots,5$. 
   [In this figure, the symbols $t_{n}$ and $\lambda_{n}$, $n=1,2,\cdots,5$ are explicitly shown only for the FTLLE $\lambda (t)$.]
   The straight shin line is a fit to the FTLLE $\lambda (t)$ in the time period $(300,t_{1}]$ by the power function $\lambda (t) = \beta_{1}t^{-\beta_{2}}$ with the fitting parameters $\beta_{j},j=1,2$.
}
\label{Fig6FinTimLya}
\end{center}
\vspfigC
\end{figure}  
%
   In Fig. \ref{Fig6FinTimLya} we show an example of time dependencies of the FTLLE $\lambda (t)$ as circles, as well as the times $t_{n}$ for the $n$-th escape of disks from the box as a vertical line ($n=1,2,\cdots,5$). 
   Here, we used the $5$-disk system whose $n$-particle survival probabilities $P_{n}(t)$ are shown in Fig. \ref{Fig2NParSurProb}.
   As expected in escape systems, the FTLLE $\lambda (t)$ decays globally, although it can increase temporally by disk-disk collisions. 
   It is important to note for contents of this paper that some local decays of $\lambda (t)$ looks like a straight line in Fig. \ref{Fig6FinTimLya} as a log-log plot of $\lambda (t)$ as a function of $t$. 
   To specify this property quantitatively, we fitted the FTLLE $\lambda (t)$ in the time period $(300,t_{1}]$ after the $(N-1)$-th disk escapes at the time $t = t_{2}$ $(< 300)$ until the $N$-th (i.e., the last) disk escapes at the time $t=t_{1}$, to a power function $\beta_{1} t^{-\beta_{2}}$ with fitting parameters $\beta_{1}$ and $\beta_{2}$, as shown the straight shin line in Fig. \ref{Fig6FinTimLya}. 
   Here, we used the values $\beta_{1} = 128$ and $\beta_{2} =  0.981$ for these fitting parameters. 
   In this time period, there is only one disk inside the box with no disk-disk collision and the system is not chaotic, so this result suggests that the FTLLE decays almost with a power law $\sim t^{-1}$ after a long time for non-chaotic cases, while the FTLLE should approach to a nonzero finite value for chaotic cases with disk-disk collisions. 
%

   For a comparison, we also show in Fig. \ref{Fig6FinTimLya} an example of time dependencies of the FTLLE $\lambda^{\prime} (t)$ of the system without any disk-disk collision as triangles.  
   Here, for the calculation of the FTLLE $\lambda^{\prime} (t)$ we used the same values of system parameters and the initial distribution of $\delta \Gamma (0)$ as those for the FTLLE $\lambda (t)$ in Fig. \ref{Fig6FinTimLya} except for absence of the disk-disk collision, and for the FTLLE $\lambda^{\prime} (t)$ the initial distribution of $\Gamma (0)$ is taken as same as the system whose $n$-particle survival probabilities $P_{n_{0}}^{\prime}(t)$ are shown in Fig. \ref{Fig4SurProbComp}. 
   Different from the FTLLE $\lambda (t)$ for the system with disk-disk collisions, the FTLLE $\lambda^{\prime} (t)$ in this figure shows a smoothly decreasing function of time, except for some abrupt changes of $\lambda^{\prime} (t)$ at escape times of disks.

\subsection{Averages and fluctuations of the finite-time largest Lyapunov exponents at escape times}
\label{StatisticalAnalysisFTLLE}

   In general, FTLLEs $\lambda (t)$ at a finite time $t$ take various values, depending on the initial conditions of the phase space vector $\bs{\Gamma}(0)$ and its deviation  $\delta \bs{\Gamma}(0)$. 
   Therefore, we investigate decay properties of FTLLEs by means of ensemble averages over an initial distribution of $\bs{\Gamma}(0)$ and  $\delta \bs{\Gamma}(0)$.

   In order to discuss such statistical properties of FTLLEs, we use the value $\lambda_{n}$ of a FTLLE $\lambda (t)$ at the time $t = t_{n}$ when the $(N-n+1)$-th particle escape from the box occurs ($n=1,2,\cdots,N$), as shown for the FTLLE $\lambda (t)$ in Fig. \ref{Fig6FinTimLya}. 
   Here, the times $t_{n}$ for these estimations of FTLLEs are given in a calculation for the $n$-particle survival probability $P_{n}(t)$. 
   In order to calculate distributions of $\lambda_{n}$ and $t_{n}$, we use the initial ensemble in which the initial phase space vector $\bs{\Gamma} (0)$ is distributed into the micro-canonical distribution with a constant energy $E$, and components of the initial Lyapunov vector $\delta \bs{\Gamma}(0)$ are chosen as the ones uniformly distributed under the constraint with a constant value of the amplitude $|\delta \bs{\Gamma}(0)|$.

   In this subsection, we pay attention mainly to the power-law decay $\lambda (t)\sim t^{-1}$ of FTLLEs, which would characterize non-chaotic dynamics of hard-disk systems. 
   For this purpose, we investigate graphs of local time averages of $\ln\lambda_{n}$ as a function of $\ln t_{n}$, which show straight lines for their power-law decays, as a useful presentation of differences between their power-law decays and non-power-law decays. 
   For local averages to obtain their smooth graphs, we use two kinds of local averages $\langle\cdots\rangle_{t}$ and $\overline{\cdots}$ for $\ca{N}_{e} = 5\times 10^{8}$ number of ensembles of $\lambda_{n} = \lambda_{n}^{(j)}$ and $t_{n} = t_{n}^{(j)}$, $j=1,2,\cdots,\ca{N}_{e}$, with $t_{n}^{(1)} \leq t_{n}^{(2)} \leq \cdots \leq t_{n}^{(\ca{N}_{e})}$. 
   The first average $\langle X(\lambda_{n},t_{n})\rangle_{t}$ of a function $X(\lambda_{n},t_{n})$ of $\lambda_{n}$ and $t_{n}$ means to take the arithmetic means of the data over every $\ca{N}_{a} = 5\times 10^{3}$ values from the beginning of the sorted group $\{X(\lambda_{n}^{(1)},t_{n}^{(1)}), X(\lambda_{n}^{(2)},t_{n}^{(2)}),\cdots,X(\lambda_{n}^{(\ca{N}_{e})},t_{n}^{(\ca{N}_{e})})$\}, so that we obtain the local averages $\langle X(\lambda_{n},t_{n})\rangle_{t}$ as $X_{n}^{(k)} \equiv \sum_{j=1+(k-1)\ca{N}_{a}}^{k\ca{N}_{a}} X(\lambda_{n}^{(j)},t_{n}^{(j)})/\ca{N}_{a}$, $k = 1,2,\cdots, \ca{N}_{e}/\ca{N}_{a}$. 
   To smooth out a graph of the data $X_{n}^{(k)}$ without reducing their data points we further take the second local average $\overline{\langle X(\lambda_{n},t_{n})\rangle_{t}}$ as $\overline{X_{n}^{(k)}} \equiv \sum_{l = k-\tilde{\ca{N}}_{k}}^{k+\tilde{\ca{N}}_{k}} X_{n}^{(l)}/(2\tilde{\ca{N}}_{k}+1)$ with $\tilde{\ca{N}}_{k} = \mbox{Min}(10^{3},(\ca{N}_{e}/\ca{N}_{a})-k,k-1)$, $k=1,2,\cdots, \ca{N}_{e}/\ca{N}_{a}$. 
   The reason to use the second average $\overline{\cdots}$ for FTLLEs is that data points of FTLLEs $\lambda_{n}$ for their power-law decays are quite a little, so that a power-law decay behavior of FTLLEs often disappears if we make a smooth graph of local averages of FTLLEs by using only the first average $\langle\cdots\rangle_{t}$ with a large number of $\ca{N}_{a}$.

\begin{figure}[!t]
\vspfigA
\begin{center}
\includegraphics[width=\widthfigC]{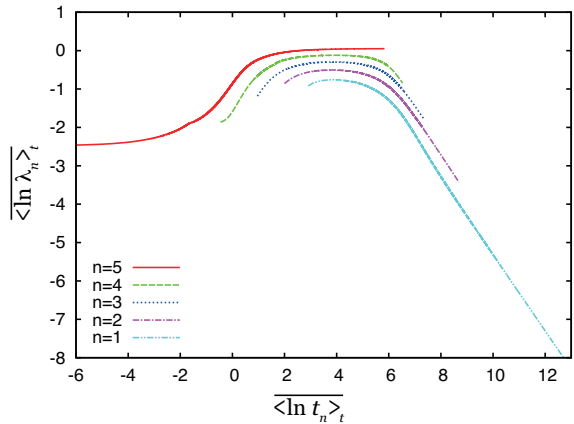}
\vspfigB
\caption{
   (Color online)
   The local time averages $\overline{\langle \ln\lambda_{n} \rangle_{t}}$ of $\ln\lambda_{n}$ as functions of local time averages $\overline{\langle \ln t_{n} \rangle_{t}}$ of $\ln t_{n}$, for $n = 5$ (the solid line), $4$ (the dashed line), $3$ (the dotted line), $2$ (the dash-dotted line) and $1$ (the dash-double-dotted line), for a $5$-disk system. 
}
\label{Fig7FtlleAve}
\end{center}
\vspfigC
\end{figure}  
%
   Figure \ref{Fig7FtlleAve} is the graphs of the local time averages $\overline{\langle \ln\lambda_{n} \rangle_{t}}$ of $\ln\lambda_{n}$ as functions of local time averages $\overline{\langle \ln t_{n} \rangle_{t}}$ of $\ln t_{n}$, for $n = 5$ (the solid line), $4$ (the dashed line), $3$ (the dotted line), $2$ (the dash-dotted line) and $1$ (the dash-double-dotted line). 
   The system is the same as that whose $n$-particle survival probabilities were shown in Fig. \ref{Fig2NParSurProb}.  
   This figure suggests that a local average of the $N$-th FTLLE $\lambda_{N}$ converges to the largest Lyapunov exponent after a long time without decay, while the other $n$-th FTLLEs $\lambda_{n}$, $n=1,2,\cdots,N-1$, decay with a power law after some times because there are quite few disk-disk collisions in orbits taking long times up to the $(N-n+1)$-th disk escape from the box, namely almost non-chaotic orbits with zero Lyapunov exponents. 
   Therefore, a transition from chaotic orbits to non-chaotic orbits would be characterized by the time starting a decay of local averages of the FTLLEs.

\begin{figure}[!t]
\vspfigA
\begin{center}
\includegraphics[width=\widthfigC]{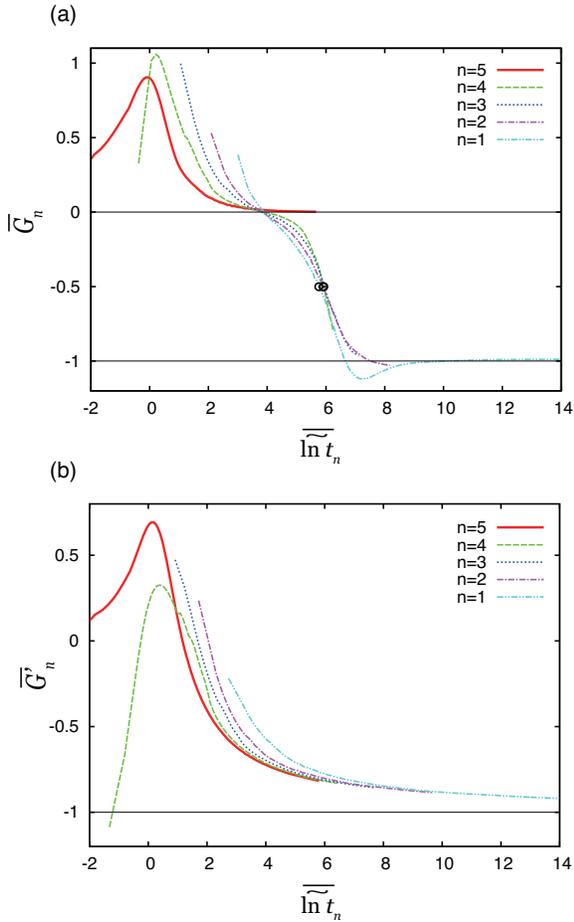}
\vspfigB
\caption{
   (Color online)
   (a) The slopes $\overline{G}_{n}$ as functions of the local average $\overline{\widetilde{\ln t}}_{n}$ of $\widetilde{\ln t}_{n}$ for $n = 5$ (the solid line), $4$ (the dashed line), $3$ (the dotted line), $2$ (the dash-dotted line) and $1$ (the dash-double-dotted line), for a $5$-disk system with disk-disk collisions. 
   Here, the open circles are the points for the slopes $\overline{G}_{n}$, $n=4,3,2,1$, to have the value $-1/2$.
   (b) The slopes $\overline{G^{\prime}}_{n}$ as functions of the local average $\overline{\widetilde{\ln t}}_{n}$ for $n = 5$ (the solid line), $4$ (the dashed line), $3$ (the dotted line), $2$ (the dash-dotted line) and $1$ (the dash-double-dotted line), for a $5$-disk system without any disk-disk collision. 
}
\label{Fig8FtlleSlop}
\end{center}
\vspfigC
\end{figure}  
%
   In order to investigate quantitatively power-law decays of local averages of the $n$-th FTLLEs $\lambda_{n}$, $n=1,2,\cdots,N-1$ after a long time, we calculate the slope $\overline{G}_{n}$ of the local averages $\overline{\langle \ln\lambda_{n} \rangle_{t}}$ of $\ln\lambda_{n}$ as functions of the local averages $\overline{\langle \ln t_{n} \rangle_{t}}$ of $\ln t_{n}$. 
   A constant value of the slope $\overline{G}_{n}$ in time suggests a power-law decay $\sim t^{\overline{G}_{n}}$ of the local average of the $n$-th FTLLEs $\lambda_{n}$.  
   In Fig. \ref{Fig8FtlleSlop}(a), we plotted the slopes as the local averages $\overline{G}_{n}$ of $G_{n} \equiv \Bigl[\overline{\langle \ln\lambda_{n} \rangle_{t}}^{(j+1)} - \overline{\langle \ln\lambda_{n} \rangle_{t}}^{(j)}\Bigr]$ $\Bigl/\Bigl[\overline{\langle \ln t_{n} \rangle_{t}}^{(j+1)} - \overline{\langle \ln t_{n} \rangle_{t}}^{(j)}\Bigr]$ as functions of the local average $\overline{\widetilde{\ln t}}_{n}$ of $\widetilde{\ln t}_{n} \equiv \Bigl[\overline{\langle \ln t_{n} \rangle_{t}}^{(j+1)} + \overline{\langle \ln t_{n} \rangle_{t}}^{(j)}\Bigr]\Bigr/2$, for $n = 5$ (the solid line), $4$ (the dashed line), $3$ (the dotted line), $2$ (the dash-dotted line) and $1$ (the dash-double-dotted line), in which $\overline{\langle \ln\lambda_{n} \rangle_{t}}^{(j)}$ and its corresponding quantity $\overline{\langle \ln t_{n} \rangle_{t}}^{(j)}$, $j=1,2,\cdots,$ $\ca{N}_{e}/\ca{N}_{a}$, are the $j$-th values of $\overline{\langle \ln\lambda_{n} \rangle_{t}}$ and $\overline{\langle \ln t_{n} \rangle_{t}}$, respectively, with  $\overline{\langle \ln t_{n} \rangle_{t}}^{(1)} < \overline{\langle \ln t_{n} \rangle_{t}}^{(2)} < \cdots < \overline{\langle \ln t_{n} \rangle_{t}}^{(\ca{N}_{e}/\ca{N}_{a})}$. 
   Figure \ref{Fig8FtlleSlop}(a) with Fig. \ref{Fig7FtlleAve} suggests that the local averages of the $n$-th FTLLEs $\lambda_{n}$, $n = 1,2,3$ decay asymptotically with a power law $\sim t^{-1}$, while the one of the $N$-th FTLLE $\lambda_{N}$ seems to converge to a positive value, i.e., the largest Lyapunov exponent, without decaying to zero. 
   For a comparison, in Fig. \ref{Fig8FtlleSlop}(b) we show the slopes $\overline{G^{\prime}}_{n}$ as functions of the local average $\overline{\widetilde{\ln t}}_{n}$ for the system without any disk-disk collision, for $n = 5$ (the solid line), $4$ (the dashed line), $3$ (the dotted line), $2$ (the dash-dotted line) and $1$ (the dash-double-dotted line). 
   Here, the slopes $\overline{G^{\prime}}_{n}$ were calculated in the same system and the same calculation process as the one for the slope $\overline{G}_{n}$ except for an absence of disk-disk collision and the initial distribution of $\Gamma (0)$ used in the system whose $n$-particle survival probabilities $P_{n_{0}}^{\prime}(t)$ are shown in Fig. \ref{Fig4SurProbComp}. 
   For the case without any disk-disk collision, all of the slopes  $\overline{G^{\prime}}_{n}$, $n = 1,2,\cdots,N$, seem to approach gradually to the value $-1$ as the time goes to infinity. 
   In contrast, the slopes $\overline{G}_{n}$, $n = 1,2,\cdots,N-1$, for the case with disk-disk collisions, show rapid decays from values around zero to the value $-1$.

\begin{figure}[!t]
\vspfigA
\begin{center}
\includegraphics[width=\widthfigC]{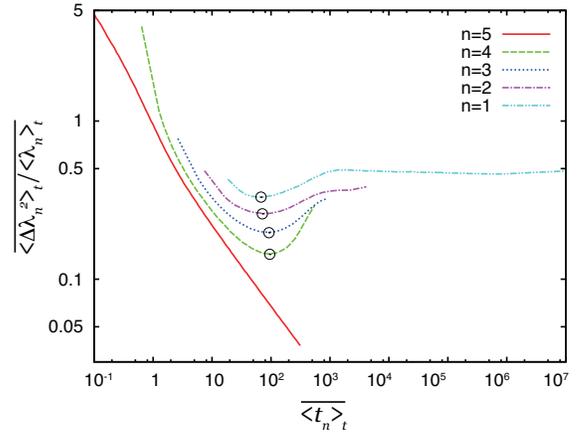}
\vspfigB
\caption{
   (Color online)
   The averaged ratios $\overline{\langle\Delta\lambda_{n}^{2}\rangle_{t}/\langle\lambda_{n}\rangle_{t}}$ as functions of the local averages $\overline{\langle t_{n}\rangle_{t}}$ of $t_{n}$, for $n = 5$ (the solid line), $4$ (the dashed line), $3$ (the dotted line), $2$ (the dash-dotted line) and $1$ (the dash-double-dotted line) for a $5$-disk system. 
   The open circles are the points at which the quantities $\overline{\langle\Delta\lambda_{n}^{2}\rangle_{t}/\langle\lambda_{n}\rangle_{t}}$, $n=4,3,2,1$ take their local minima. 
}
\label{Fig9FtlleFlu}
\end{center}
\vspfigC
\end{figure}  
%
   The FTLLEs $\lambda_{n}$ as functions of the escape times $t_{n}$ for $n = 1,2,\cdots,N$ have large fluctuations, so that it would be meaningful to discuss not only their averages but also their fluctuations.  
   As a feature of fluctuations of the FTLLEs $\lambda_{n}$, in Fig. \ref{Fig9FtlleFlu} we show the graphs of the averaged ratios $\overline{\langle\Delta\lambda_{n}^{2}\rangle_{t}/\langle\lambda_{n}\rangle_{t}}$ between the local time averages $\langle\Delta\lambda_{n}^{2}\rangle_{t}$ of the variances $\Delta\lambda_{n}^{2} \equiv (\lambda_{n} - \langle\lambda_{n} \rangle_{t} )^{2}$ of $\lambda_{n}$ and the local time averages $\langle\lambda_{n}\rangle_{t}$ of $\lambda_{n}$, as functions of local averages $\overline{\langle t_{n}\rangle_{t}}$ of $t_{n}$ , for $n = 5$ (the solid line), $4$ (the dashed line), $3$ (the dotted line), $2$ (the dash-dotted line) and $1$ (the dash-double-dotted line), in the same system as that whose $n$-particle survival probabilities were shown in Fig. \ref{Fig2NParSurProb}. 
   This figure suggests that a fluctuation amplitude of the $N$-th FTLLE $\lambda_{N}$ decreases monotonically as a function of time. 
   This is consistent with the fact that disks do not escape until the time $t_{N}$ and the FTLLEs for the system without any disk escape converges to a (positive) Lyapunov exponent as time goes to infinity for almost any initial conditions, known as Oseledec's theorem \cite{S91}. 
   In contrast, it is shown in Fig. \ref{Fig9FtlleFlu} that variances of the $n$-th FTLLEs $\lambda_{n}$, divided by the local averages $\langle \lambda_{n} \rangle_{t}$ of the FTLLEs $\lambda_{n}$, for $n=1,2,\cdots,N-1$, have local minima (the open circles in Fig. \ref{Fig9FtlleFlu}) as functions of time. 
   After their local minima, the ratios $\overline{\langle\Delta\lambda_{n}^{2}\rangle_{t}/\langle\lambda_{n}\rangle_{t}}$, $n=1,2,\cdots,N-1$, increase in time, then seem to reach almost to a constant value as shown for $n=1,2$  in Fig. \ref{Fig9FtlleFlu}, meaning that in this time period variances of the $n$-th FTLLEs $\lambda_{n}$ keep to have a similar amplitude to the local average of the FTLLEs which would go to zero as time goes to infinity.  
   These increases of the quantities $\overline{\langle\Delta\lambda_{n}^{2}\rangle_{t}/\langle\lambda_{n}\rangle_{t}}$ as function of time are supposed to come from non-chaotic orbits which take long times for the $(N-n+1)$-th disk escape from the box.

\subsection{Relations among transition times in decays of the survival probabilities and the finite-time largest Lyapunov exponents}

   Now, we came to the stage of discussions on a direct connection between decays of survival probabilities (characterizing escape properties) and FTLLEs (characterizing chaotic or non-chaotic properties) for many-hard-disk systems. 
   For such discussions we introduce three different types of times to characterize decay transitions of the survival probabilities or the FTLLEs as follows.

   First, as discussed in Secs. \ref{ExponentialDecaysSurvivalProbabilities} and \ref{PowerDecaysSurvivalProbabilities}, decays of the $n$-particle survival probabilities $P_{n}(t)$, $n=2,3,\cdots,N-1$, transfer from the superpositions (\ref{SurviProbaExpon1}) of exponential decays to the power-law decays (\ref{PowerDecaySurProbCol1}) for many-hard-disk systems. 
   In order to represent quantitatively intermediate times between their decays (\ref{SurviProbaExpon1}) and  (\ref{PowerDecaySurProbCol1}), we introduce the times $t_{n}^{(sur)}$ as the ones for the slope $F_{n}(t)$ to cross the line $-(2n-1)$ for $n=2,3,\cdots,N-1$. 
   We also introduce the time $t_{1}^{(sur)}$ as the one for the slope $F_{1}(t)$ to cross the line $-1/2$. 
   The survival probabilities at these times are indicated by the open circles in Fig. \ref{Fig5SlopSurProb}(a).

   Second, as discussed in Sec. \ref{StatisticalAnalysisFTLLE}, the local averages of FTLLEs $\lambda_{n}$ at the escape times $t_{n}$, $n=1,2,\cdots,N-1$, show a power-law decay $\sim t^{-1}$ after a long time. 
   These power-law decays would be caused by non-chaotic orbits of disks which would have longer escape times than those of chaotic orbits with frequent disk-disk collisions. 
   We introduce the times $t_{n}^{(lya,ave)}$, $n=1,2,\cdots,N-1$, as those for the slopes $\overline{G}_{n}$, $n=1,2,\cdots,N-1$, to take the value $-1/2$, respectively, and we use these times to estimate the times after which orbits are almost non-chaotic. 
   The local-averaged FTLLEs at the times $t_{n}^{(lya,ave)}$, $n=1,2,\cdots,N-1$, are indicated by the open circles in Fig. \ref{Fig8FtlleSlop}(a).

   The third type of times to characterize a dynamical transition is related to fluctuation properties of FTLLEs discussed in Sec. \ref{StatisticalAnalysisFTLLE}. 
   These times, represented as $t_{n}^{(lya,flu)}$, $n=1,2,\cdots,N-1$ in this paper, are defined by the times when the averaged ratios $\overline{\langle\Delta\lambda_{n}^{2}\rangle_{t}/\langle\lambda_{n}\rangle_{t}}$, $n=1,2,\cdots,N-1$, take their local minima, respectively.
   The averaged ratios $\overline{\langle\Delta\lambda_{n}^{2}\rangle_{t}/\langle\lambda_{n}\rangle_{t}}$ at these times are indicated by the open circles in Fig. \ref{Fig9FtlleFlu}.

\begin{figure}[!t]
\vspfigA
\begin{center}
\includegraphics[width=\widthfigC]{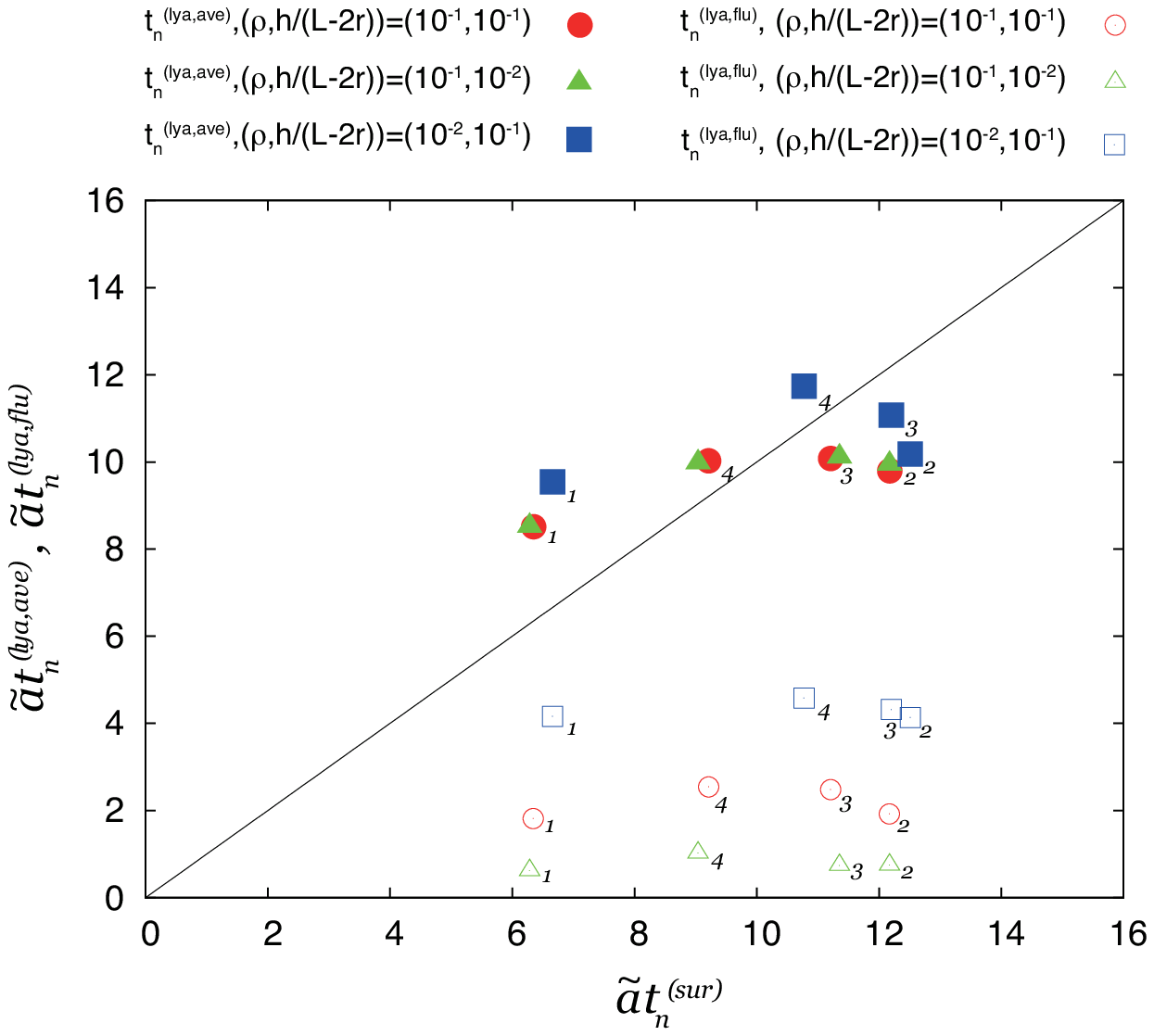}
\vspfigB
\caption{
   (Color online)
   The points $\left(\tilde{a}t_{n}^{(sur)},\tilde{a}t_{n}^{(lya,ave)}\right)$, $n=1,2,3,4$, for $\left(\rho,h/(L-2r)\right) = \left(10^{-1},10^{-1}\right)$ (the closed circles), $\left(\rho,h/(L-2r)\right) = \left(10^{-1},10^{-2}\right)$ (the closed triangles), $\left(\rho,h/(L-2r)\right) = \left(10^{-2},10^{-1}\right)$ (the closed squares),  and the points $\left(\tilde{a}t_{n}^{(sur)},\tilde{a}t_{n}^{(lya,flu)}\right)$, $n=1,2,3,4$ for $\left(\rho,h/(L-2r)\right) = \left(10^{-1},10^{-1}\right)$ (the open circles), $\left(\rho,h/(L-2r)\right) = \left(10^{-1},10^{-2}\right)$ (the open triangles), $\left(\rho,h/(L-2r)\right) = \left(10^{-2},10^{-1}\right)$ (the open squares) for $5$-disk systems. 
   The index number at the right-low side of each point corresponds to the value $n$ of each data for its point. 
   The straight line shows the cases in which $\tilde{a}t_{n}^{(sur)}$ is equal to $\tilde{a}t_{n}^{(lya,ave)}$ or $\tilde{a}t_{n}^{(lya,flu)}$.
}
\label{Fig10RelSurLya}
\end{center}
\vspfigC
\end{figure}  
%
   In order to discuss relations among these times $t_{n}^{(sur)}$,  $t_{n}^{(lya,ave)}$ and $t_{n}^{(lya,flu)}$, we show in Fig. \ref{Fig10RelSurLya} the points $\left(\tilde{a}t_{n}^{(sur)},\tilde{a}t_{n}^{(lya,ave)}\right)$, $n=1,2,3,4$ for $\left(\rho,h/(L-2r)\right) = \left(10^{-1},10^{-1}\right)$ (the closed circles), $\left(\rho,h/(L-2r)\right) = \left(10^{-1},10^{-2}\right)$ (the closed triangles), $\left(\rho,h/(L-2r)\right) = \left(10^{-2},10^{-1}\right)$ (the closed squares),  and the points $\left(\tilde{a}t_{n}^{(sur)},\tilde{a}t_{n}^{(lya,flu)}\right)$, $n=1,2,3,4$ for $\left(\rho,h/(L-2r)\right) = \left(10^{-1},10^{-1}\right)$ (the open circles), $\left(\rho,h/(L-2r)\right) = \left(10^{-1},10^{-2}\right)$ (the open triangles), $\left(\rho,h/(L-2r)\right) = \left(10^{-2},10^{-1}\right)$ (the open squares), for the $5$-disk systems whose system parameters are the same as the corresponding ones used in Table \ref{ExponDEcayFitting}. 
   Here, the index number at the right-low side of each point indicates the value $n$ of each data for its point, and the factor $\tilde{a} = hv_{0}/(L-2r)^{2}$ is the same as that used in Table \ref{ExponDEcayFitting}.

   It is shown in Fig. \ref{Fig10RelSurLya} that the time $t_{n}^{(sur)}$ takes a quite similar value to the time $t_{n}^{(lya, ave)}$, $n=1,2,\cdots,N-1$, for different particle densities $\rho$ and hole size ratios $h/(L-2r)$, noting that the points $\left(\tilde{a}t_{n}^{(sur)},\tilde{a}t_{n}^{(lya,ave)}\right)$ are close on the straight line to indicate the relation $\tilde{a}t_{n}^{(sur)}=\tilde{a}t_{n}^{(lya,ave)}$. 
   This result gives a supportive evidence that a transition from a decay represented as a superposition (\ref{SurviProbaExpon1}) of exponential decays to a power-law decay (\ref{PowerDecaySurProbCol1}) for the $n$-particle survival probability $P_{n}(t)$ corresponds to a transition of the FTLLE $\lambda_{n}$ to its power-law decay caused by non-chaotic properties of the system. 
   In other words, the origin of power-law decays of the $n$-particle survival probabilities would be non-chaotic orbits of many hard disks. 
   It should also be noted that the points $\left(t_{n}^{(sur)},t_{n}^{(lya,ave)}\right)$ are well-scaled by multiplying the factor $\tilde{a}$, and especially the points $\left(t_{n}^{(sur)},t_{n}^{(lya,ave)}\right)$ for different hole sizes with $h/(L-2r) = 10^{-1}$ and $10^{-2}$ almost coincide in Fig. \ref{Fig10RelSurLya}. 
   On the other hand, we could not find such kinds of equivalence and scaling between the time $t_{n}^{(sur)}$ and $t_{n}^{(lya,flu)}$, although $\tilde{a}t_{n}^{(sur)}$ and $\tilde{a}t_{n}^{(lya,flu)}$ have similar orders of magnitudes, as shown in Fig. \ref{Fig10RelSurLya}. 
   This figure also shows that the quantities $\tilde{a}t_{n}^{(lya,flu)}$, $n=1,2,\cdots,N-1$, decrease as the hole size ratio $h/(L-2r)$ decreases, while they increase as the initial particle density $\rho$ decreases.

\section{Conclusions and remarks} 
\label{ConclusionRemarks}

   In this paper, we have discussed escape properties of many hard disks from a square box via a hole. 
   To investigate disk escapes, the $n$-particle survival probability $P_{n}(t)$ was introduced as the probability for $n\;(\leq N)$ disks to remain inside the box without escaping up to the time $t \; (\geq 0)$, starting from $N$ disks inside the box at the initial time $t=0$. 
   At early times the probabilities $P_{n}(t), n=2,3,\cdots,N-1$, decay as superpositions of exponential functions, while after a long time the probabilities $P_{n}(t), n=1,2,\cdots,N-1$, show power-law decays. 
   Especially, as an important effect of disk-disk collisions, the exponent of the power-law decay for the probability $P_{n}(t)$ after a long time is given by $-n$ for $n=1$ and $-2n$ for $n=2,3,\cdots$ [as shown in Eq. (\ref{PowerDecaySurProbCol1})], in contrast to the case without any disk-disk collision in which the exponent of the power-law decay for the $n$-particle survival probability $P_{n}^{\prime}(t)$ after a long time is simply given by $-n$ for $n=1,2,\cdots$ [as shown in Eq. (\ref{PowerDecaySurProbNoCol1})]. 
   These power-law decays of the $n$-particle survival probabilities after a long time for many-hard-disk systems can be verified in various particle densities, hole sizes, particle numbers, including the cases shown in Table \ref{ExponDEcayFitting}.
   This suggests a possibility to obtain informations on the particle number inside the box and particle-particle interactions from decay behaviors of the $n$-particle survival probabilities.  
   We also discussed scaling properties for exponential decay rates of the probabilities $P_{n}(t)$ for various hole sizes and initial particle densities.

   In order to discuss escape features based on dynamical characteristics of many-particle systems, we further discussed properties of the finite time largest Lyapunov exponents (FTLLEs) in escape systems consisting of many hard disks. 
   The FTLLE is defined as the exponential rate of expansion or contraction of the absolute magnitude of a small deviation of the phase space vector of the system at a finite time, and it converges to a non-zero finite value as the largest Lyapunov exponent for chaotic systems with a dynamical instability in the long time limit. 
   The dynamics of an escape system consisting of many hard disks in a box should be chaotic at early times for disk-disk collisions, but it becomes non-chaotic in the long time limit because only a single (or even zero) disk exists inside the box after a long time when other disks have already escaped from the box via a hole. 
   In this sense, a transition from a chaotic dynamics to non-chaotic dynamics occurs in such escape systems consisting of many hard disks. 
   In this paper, this dynamical transition was discussed by decay behaviors of FTLLEs of the escape systems. 
   We introduced the FTLLE $\lambda_{n}$ of a many-hard-disk system in a box with a hole at the time $t=t_{n}$ when the $(N-n+1)$-th disk escape from the box via the hole occurs. 
   It was shown that local time averages of the FTLLEs $\lambda_{n}$, $n=1,2,\cdots,N-1$, decay with a power law $\sim t^{-1}$ after a long time, suggesting that the transition from a chaotic dynamics to a non-chaotic dynamics in escape systems could be described by the transition from the value $0$ of the slope of a local time average of $\ln \lambda_{n}$ with respect to a local time average of $\ln t_{n}$ to its value $-1$. 
   We estimated this transition time as the time when this slope of the local time average of $\ln \lambda_{n}$ takes the intermediate value $-1/2$, and showed that this time could be strongly correlated to the transition time for a power-law decay of the $n$-particle survival probability. 
   This result would give a quantitative evidence on a connection between the escape features of many-hard-disk systems and their chaotic or non-chaotic dynamical characteristics. 
   It was also shown that these transition times show scaling behaviors on initial particle densities and hole size ratios. 
   We also discussed a fluctuation property of the FTLLEs $\lambda_{n}$ in the sense that a local time average of the variation of $\lambda_{n}$ divided by a local time average of $\lambda_{n}$ takes a local minimum, then it would take an almost constant value locally in time for $n=1,2,\cdots,N-1$, as a function of time. 

\begin{figure}[!t]
\vspfigA
\begin{center}
\includegraphics[width=\widthfigC]{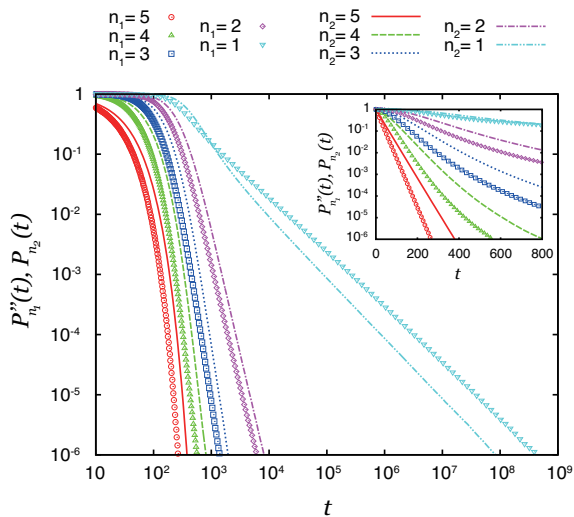}
\vspfigB
\caption{
   (Color online)
   The $n_{1}$-particle survival probabilities $P_{n_{1}}^{\prime\prime}(t)$ of a $5$-disk system as functions of time $t$ for $n_{1} = 5$ (the circles), $4$ (the triangles), $3$ (the squares), $2$ (the diamonds) and $1$ (the inverted triangles) for the hole of the length $r+h$ in a {\em single side} from a single corner of the box; 
and the $n_{2}$-particle survival probabilities $P_{n_{2}}(t)$ of a $5$-disk system as functions of time $t$ for $n_{2} = 5$ (the solid line), $4$ (the dashed line), $3$ (the dotted line), $2$ (the dash-dotted line), $1$ (the dash-double-dotted line) for the hole of the length $r+(h/2)$ in {\em both the sides} from a single corner of the box. 
   The main figure and the inset are for graphs of $P_{n_{1}}^{''}(t)$ and $P_{n_{2}}(t)$ on log-log plots and linear-log plots, respectively.
}
\label{Fig11SurProbHole}
\end{center}
\vspfigC
\end{figure}  
%
%
   In the escape models discussed in this paper, we put a single hole for disks to escape, which consists of two regions over {\em two sides} from a single corner of the box. 
   This hole configuration is chosen so that orbit properties for a disk bouncing between two confronting walls without any collision with other disks, the so-called bouncing ball (or sticky) orbits,  are the same for two kinds of the confronting walls. 
   The bouncing ball orbits are known to play an essential role in power-law decays of survival probabilities in escapes of single-particle two-dimensional billiard models \cite{AH04,DG09}. 
   On the other hand, it would be meaningful to note what happens in particle escapes for different hole configurations. 
   As such an example, in Fig. \ref{Fig11SurProbHole} we show the  $n_{1}$-particle survival probabilities $P_{n_{1}}^{\prime\prime}(t)$ of a $5$-disk system as functions of time $t$ for $n_{1} = 5$ (the circles), $4$ (the triangles), $3$ (the squares), $2$ (the diamonds) and $1$ (the inverted triangles)  for the hole of the length $r+h$ in a {\em single side} from a single corner of the box. 
   Here, except for the hole position we used the same values of the system parameters and the same initial distributions as the system whose survival probabilities are shown in Fig. \ref{Fig2NParSurProb}.  
   [In Fig. \ref{Fig11SurProbHole}, for comparisons with the survival probabilities $P_{n_{1}}^{\prime\prime}(t)$, we also show graphs of the $n_{2}$-particle survival probabilities $P_{n_{2}}(t)$ for $n_{2} = 5$ (the solid line), $4$ (the dashed line), $3$ (the dotted line), $2$ (the dash-dotted line), $1$ (the dash-double-dotted line), which are the same as in Fig. \ref{Fig2NParSurProb}.] 
   Figure \ref{Fig11SurProbHole} shows that the $n$-particle survival probabilities $P_{n}^{\prime\prime}(t)$ of the escape system with the hole in the single side of the box decay faster than the $n$-particle survival probabilities $P_{n}(t)$ of the system with the hole in both the sides of box for $n = 2,3,\cdots,N$, while the survival probability $P_{1}^{\prime\prime}(t)$ decays slower (faster) than the survival probability $P_{1}(t)$ after a long time (at early times). 
   It should be also noted that the power-law decay $\sim t^{-1}$ of the survival probability $P_{1}^{\prime\prime}(t)$ after a long time looks to be much weaker than that of $P_{1}(t)$.  
   In general, escape behaviors of disks from the box could depend on where we put a hole, because of not only the properties of bouncing ball orbits but also finite size effects of disks \cite{MemoHole}. 
   Another problem involving finite size effects of disks would be an accurate evaluation of decay rates of survival probabilities based on the ergodicity of many-hard-disk systems.

   The $n$-particle survival probability is given from times for one of a finite number of disks to escape from the box in many ensembles of orbits. 
   In this sense it could be discussed as escapes of a single particle in a high dimensional coordinate space.  
   Recently, escape behaviors of a single particle in a high dimensional chaotic system, such as asymptotic decays of a survival probability in the high dimensional Lorenz gas model, have been discussed \cite{D12,NS14}. 
   As another related topic on dynamical decays of many-particle systems, an asymptotic power-law decay $\sim t^{1-\kappa}$ of the probability for $\kappa$ particles of not colliding before time $t$ with periodic boundary conditions was discussed in Ref. \cite{D14}. 
    In these works, a special type of particle orbits without any collision with other disks plays an essential role in asymptotic decays of survival probabilities, etc. 
    On the other hand, the difference between Eqs. (\ref{PowerDecaySurProbCol1}) and (\ref{PowerDecaySurProbNoCol1}) suggests that for many hard disks in a box with a hole, disk-disk collisions are not negligible even in orbits with long escape times. 
    Actually, we can check, for example, that even in the orbit with the longest escape time $t_{1}$ in the $5\times 10^{8}$ ensembles of our numerical calculations for the escape system whose $n$-particle survival probabilities are represented in Fig. \ref{Fig2NParSurProb}, there are dozens of disk-disk collisions before all disks to escape from the box.
    Therefore, it would be an important future problem to clarify what types of orbits dominate asymptotic power-law decays of $n$-particle survival probabilities in many-hard-disk systems.

   As a dynamical description of escape phenomena in chaotic systems with a single particle, the escape rate formalism is known \cite{GN90,G98,D99,K07}. 
   This formalism describes the escape dynamics by the repeller \cite{D99,CRM10}, which is introduced as the set of phase space points remaining inside an initially confined region forever in escape systems. 
   By its definition, a particle on a repeller never leave the initially confined region, and finite time Lyapunov exponents of the particle on the repeller can approach non-zero values (as Lyapunov exponents) for chaotic systems in the long time limit. 
   Based on these Lyapunov exponents, the escape rate formalism shows that a difference between the sum over the positive Lyapunov exponents and the Kolmogorov-Sinai entropy on the repeller gives an escape rate of the exponential decay of a survival probability of one-particle escape systems.  
   In contrast, in this paper we discussed finite time Lyapunov exponents for orbits of escaping particles. 
   In these orbits any particle escapes from the box at a finite time, so that finite time Lyapunov exponents for these orbits of particles inside an initially confined region go to zero at a finite time, different from the Lyapunov exponents on the repeller. 
   There could exist repellers in many-particle systems, but it is very difficult to analyze them analytically or numerically at the present time. 
   It would be interesting to discuss how we apply dynamical theories for the repeller of escape systems, such as the escape rate formalism, to many-particle systems or non-exponential decays of survival probabilities.

   Dynamical features on many-particle effects in escape phenomena have been discussed recently in quantum systems by using the survival probabilities for all particles to remain inside a finite region \cite{TS11a,C11,GGL12}.  
   It was shown in these works that the survival probabilities in quantum systems with non-interacting and identical many particles show power-law decays asymptotically in time, and exponents of their power decays depend on the quantum statistics, i.e., whether the particles are identical fermions or bosons. 
   Many interesting features, such as effects of various particle-particle interactions or many holes and quantum-classical correspondences, etc., would remain as open problems on escape phenomena of many-particle systems. 



\appendix
\section{Derivations of Eqs. (\ref{EscapTimeProba3b}) and (\ref{SurviProbaExpon1})} 
\label{LaplaceHeaviside}

   In this appendix, we derive Eq. (\ref{EscapTimeProba3b}) from Eqs. (\ref{FunctTildeF1}) and (\ref{EscapTimeProba3a}), as well as Eq. (\ref{SurviProbaExpon1}) from Eqs. (\ref{EscapTimeProba1}) and (\ref{EscapTimeProba3b}).

   First, we introduce the Laplace transformation $\hat{f}_{1}(\omega;a)\equiv \int_{0}^{+\infty} d\tau\; \tilde{f}_{1}(\tau;a) \exp (-\omega \tau)$ of the function $\tilde{f}_{1}(\tau;a)$ of $\tau\; (>0)$, which is given by
\begin{eqnarray}
   \hat{f}_{1}(\omega;a) = \frac{a}{\omega + a}
\label{LaplaFunctTildeF1}
\end{eqnarray}
for $\omega + a > 0$, by using Eq. (\ref{FunctTildeF1}). 
   Second, using Eq. (\ref{LaplaFunctTildeF1}) and the convolution formula of Laplace transformations, the Laplace transformation $\hat{f}_{k}(\omega)\equiv \int_{0}^{+\infty} d\tau\; \tilde{f}_{k}(\tau) \exp (-\omega \tau)$ of the function (\ref{EscapTimeProba3a}) for $k\geq 2$ is given by
\begin{eqnarray}
   \hat{f}_{k}(\omega) &=& \prod_{j=1}^{k}  \hat{f}_{1}(\omega;a_{j}) 
      \label{LaplaFunctFn1} \\
   &=& \sum_{j=1}^{k} \frac{A_{j}}{\omega + a_{j}} .
\label{LaplaFunctFn}
\end{eqnarray}
in a form of the partial fraction under the assumption $a_{j} \neq a_{k}$ for $j\neq k$. 
   Here,  $A_{j}$ is a constant and is given by
\begin{widetext}
\begin{eqnarray}
   A_{j} &=& \lim_{\omega \rightarrow - a_{j}} \left(\omega + a_{j}\right) 
         \sum_{l=1}^{k} \frac{A_{l}}{\omega + a_{l}} 
      = \lim_{\omega \rightarrow - a_{j}} \left(\omega + a_{j}\right) 
         \prod_{l=1}^{k}  \hat{f}_{1}(\omega;a_{l}) 
      \nonumber \\
   &=& \frac{\prod_{l=1}^{k} a_{l}}{(a_{1}-a_{j})(a_{2}-a_{j})
      \cdots (a_{j-1}-a_{j})(a_{j+1}-a_{j})(a_{j+2}-a_{j})\cdots (a_{k}-a_{j})} .
\label{LaplaFunctFnCoeff1}
\end{eqnarray}
\end{widetext}
%
   From Eq. (\ref{LaplaFunctFn}), the inverse-transformation $\tilde{f}_{k}(\tau)$ of the function $ \hat{f}_{k}(\omega)$ is represented as
\begin{eqnarray}
   \tilde{f}_{k}(\tau) = \sum_{j=1}^{k} A_{j} e^{-a_{j} \tau} .
\label{EscapTimeProba3c}
\end{eqnarray}
By inserting Eq. (\ref{LaplaFunctFnCoeff1}) into Eq. (\ref{EscapTimeProba3c}), we obtain Eq. (\ref{EscapTimeProba3b}).

   Using Eqs. (\ref{LaplaFunctFn1}), (\ref{LaplaFunctFn}) and the normalization condition $\int_{0}^{+\infty} d\tau\; \tilde{f}_{1} (\tau;a)  = \hat{f}_{1}(0;a) = 1$ of the probability density $\tilde{f}_{1}(\tau;a)$, we obtain
\begin{eqnarray}
   \int_{0}^{+\infty} d\tau\; \tilde{f}_{k}(\tau) &=& \hat{f}_{k}(0) = \prod_{j=1}^{k}\hat{f}_{1}(0;a_{j}) 
      \nonumber \\
   &=& 
   1 = \sum_{j=1}^{k} \frac{A_{j}}{a_{j}}, 
\label{NormaProbaFExp1}
\end{eqnarray}
which includes the normalization condition of the probability density $\tilde{f}_{k}(\tau)$. 
   From Eqs. (\ref{EscapTimeProba1}),  (\ref{EscapTimeProba3c}) and (\ref{NormaProbaFExp1}) we derive 
\begin{eqnarray}
   \tilde{P}_{N-k+1}(t) &=& 1 - \sum_{j=1}^{k} A_{j} \frac{1-e^{-a_{j}t}}{a_{j}}
      \nonumber \\
   &=& \sum_{j=1}^{k} \frac{A_{j}}{a_{j}}  e^{-a_{j}t} 
\label{SurviProbaExpon0}
\end{eqnarray}
for the survival probability $ P_{N-k+1}(t) =  \tilde{P}_{N-k+1}(t)$ in the case of the probability density $f_{k}(\tau) = \tilde{f}_{k}(\tau)$ of the escape time $\tau$ of the $k$ disks. 
   By inserting Eq. (\ref{LaplaFunctFnCoeff1}) into Eq. (\ref{SurviProbaExpon0}), we obtain Eq. (\ref{SurviProbaExpon1}).




\begin{thebibliography}{99}
\bibitem{NoteB} 
   In a practical viewpoint, the $n$-particle survival probability $P_{n}(t)$ can be calculated as follows. 
   We prepare $\ca{M}$ number of initial ensembles of $N$ disks, and for each initial ensemble we calculate the time $t_{n}^{(j)}$ at which the $(N-n+1)$-th escape of disks via the hole occurs ($j=1,2,\cdots, \ca{M}$). 
   Then, we take the set of these times with the increasing order as $t_{n}^{(1)} \leq t_{n}^{(2)} \leq \cdots \leq t_{n}^{(\ca{M})}$. 
   In the large limit of the ensemble number $\ca{M}\rightarrow +\infty$, the quantity $\lim_{\ca{M}\rightarrow +\infty} (\ca{M} - j)/\ca{M}$ gives the $n$-particle survival probability $P_{n}(t)$ at the time $t=t_{n}^{(j)}$.
\bibitem{NoteC} In a numerical calculation of the slope of a function, the numerical data showing the function have to be enough smooth. In the actual numerical calculations of the slopes $F_{n}(t)$ of $\ln P_{n}(t)$  and  $F_{n}^{\prime}(t)$ of $\ln P_{n}^{\prime}(t)$ shown in Fig. \ref{Fig5SlopSurProb} we first took a local time average of $\ln P_{n}(t)$ and $\ln P_{n}^{\prime}(t)$, respectively, then calculated the slopes of those functions smoothed locally in time.
\end{thebibliography}
\end{document}